\documentclass[11pt]{article}
\usepackage{latexsym,amsmath,amscd,amssymb,graphicx,amsbsy,color,xcolor,amsfonts,amsthm}

\begin{document}

\title{\huge{Making the Relativistic Dynamics Equation \emph{Covariant}:}\\\Large{Explicit Solutions for Motion under a Constant Force}\thanks{Supported in part by the German-Israel Foundation for Scientific Research and Development: GIF No. 1078-107.14/2009}}
\author{Yaakov Friedman and Tzvi Scarr\\
Jerusalem College of Technology\\
Departments of Mathematics and Physics \\
P.O.B. 16031 Jerusalem 91160, Israel\\
e-mail: friedman@jct.ac.il, tzviscarr@gmail.com\\
  \\
Physica Scripta,\textbf{ 86} (2012) 065008}

\date{}
\maketitle

\begin{abstract}
We derive a 4D \emph{covariant} Relativistic Dynamics Equation. This equation canonically extends the 3D relativistic dynamics equation $\mathbf{F}=\frac{d\mathbf{p}}{dt}$, where $\mathbf{F}$ is the 3D force and $\mathbf{p}=m_0\gamma\mathbf{v}$ is the 3D relativistic momentum. The \emph{standard} 4D equation $F=\frac{dp}{d\tau}$ is only partially covariant. To achieve full Lorentz covariance, we replace the four-force $F$ by a rank 2 antisymmetric \emph{tensor} acting on the four-velocity. By taking this tensor to be constant, we obtain a covariant definition of \emph{uniformly accelerated motion}. This solves a problem of Einstein and Planck.

We compute \emph{explicit} solutions for uniformly accelerated motion. The solutions are divided into four \emph{Lorentz-invariant} types: null, linear, rotational, and general.  For \emph{null} acceleration, the worldline is \emph{cubic} in the time. \emph{Linear} acceleration covariantly extends 1D hyperbolic motion, while \emph{rotational} acceleration  covariantly extends pure rotational motion.



\end{abstract}

\newpage
\section{Introduction}\label{Intro}

$\;\;\;$

The standard 3D relativistic dynamics equation \cite{gps,Rindler,Moller}
\begin{equation}\label{dudtg}
\mathbf{F}=\frac{d\mathbf{p}}{dt}
\end{equation} is an adaptation of Newton's Second Law $\mathbf{F}=m\mathbf{a}$ to special relativity. Here, $\mathbf{F}$ is the 3D force, and $\mathbf{p}=m_0\gamma\mathbf{v}$ is the 3D relativistic momentum. When the 3D force $\mathbf{F}$ is \emph{constant}, the solutions to (\ref{dudtg}) are traditionally called \emph{uniformly accelerated motion}. Equation (\ref{dudtg}), however, is covariant only with respect to the \emph{little Lorentz group} and \emph{not} covariant with respect to the full Lorentz group.

As a 4D covariant extension of (\ref{dudtg}), we have
\begin{equation}\label{4drde}
F=\frac{dp}{d\tau},
\end{equation}
where $F$ is the four-force, $p$ is the four-momentum, and $\tau$ is proper time (see \cite{gps}). Unfortunately, equation (\ref{4drde}) is also covariant only with respect to the little Lorentz group. Moreover, when $F$ is a constant, as in a homogeneous gravitational field, equation (\ref{4drde}) has no solution! This follows from the fact that the four-velocity and the four-acceleration are perpendicular. This was noticed by Planck, who wrote to Einstein about it. This, in turn, prompted Einstein to submit a ``correction" \cite{E1} to \cite{E2}. In the correction, he states that the ``concept `uniformly accelerated' needs further clarification." This was a call for a fully Lorentz covariant relativistic dynamics equation and for a better definition of ``uniform acceleration."

It was clear, even in 1908, that the \emph{physical} definition of ``uniformly accelerated motion" is motion whose acceleration is \emph{constant in the comoving frame}. This definition is found widely in the literature, as early as \cite{Mink} and \cite{Born2}, again in \cite{LL}, and as recently as \cite{Rohrlich2} and \cite{Lyle}. This definition is natural, since the acceleration in the comoving frame is ``precisely the push we feel when sitting in an accelerating rocket" or automobile. Similarly, ``by the equivalence principle, the gravitational field in our terrestrial lab is the negative of our proper acceleration, our instantaneous rest frame being an imagined Einstein cabin falling with acceleration $g$" (\cite{Rindler}, page 71).

If the acceleration is constant in the comoving frame, then the length of the four-acceleration $a$ is constant:
\begin{equation}\label{a2isconst}
a^\mu a_\mu= \hbox{constant}.
\end{equation}
Equation (\ref{a2isconst}) is a good candidate to replace (\ref{4drde}).  It's even \emph{fully} Lorentz covariant. However, as in the case of equations (\ref{dudtg}) and (\ref{4drde}), existing techniques have produced only \emph{1D hyperbolic motion} as solutions to (\ref{a2isconst}). There are clearly some missing solutions, since equation (\ref{a2isconst}) is covariant, while the class of 1D hyperbolic motions is \emph{not}.

We are thus faced with two problems:
\vskip0.2cm
\begin{itemize}
\item[(1)] Can $\mathbf{F}=\frac{d\mathbf{p}}{dt}$ be extended to a 4D Lorentz covariant version?
\item[(2)] What is the ``right" equation for uniformly accelerated motion?
\end{itemize}
In this paper, we derive a 4D Lorentz covariant Relativistic Dynamics Equation:
\begin{equation}\label{uam1}
c\frac{du^{\mu}}{d\tau}=A^{\mu}_{\nu}u^{\nu},
\end{equation}
where $u$ is the four-velocity, $\tau$ is proper time, and $A_{\mu\nu}$ is a rank 2 antisymmetric tensor, or, equivalently, $A^\mu_\nu$ is skew adjoint with respect to the Minkowski inner product $\eta_{\mu\nu}=\operatorname{diag}(1,-1,-1,-1)$. As will be shown here, equation (\ref{uam1}) has the following advantages:
\begin{itemize}
\item [$\bullet$] It canonically extends the Relativistic Dynamics Equation (\ref{dudtg}) and is covariant with respect to the \emph{full Lorentz group}
\item [$\bullet$] By redefining \emph{uniformly accelerated motion} as the solutions to (\ref{uam1}) when $A$ is \emph{constant}, we obtain the clarification that Einstein was looking for
\item[$\bullet$] It admits four Lorentz-invariant classes of solutions: null acceleration, linear acceleration, rotational acceleration, and general acceleration. The null, rotational, and general classes were previously unknown. The linear class is a \emph{covariant extension} of 1D hyperbolic motion and contains the motion of an object in a \emph{homogeneous gravitational field}
\item[$\bullet$] It can be extended in a straightforward manner to obtain a covariant definition of the ``comoving frame" of a uniformly accelerated observer. We have shown in \cite{FS2} that all of the solutions of (\ref{uam1}) have constant acceleration in the comoving frame.
\item[$\bullet$] It can be modified to accommodate a universal maximal acceleration. Thus, this paper is an important step in the study of evidences for and implications of the existence of a universal maximal acceleration (see \cite{FG10,F11Ann,FR}).

\end{itemize}


The plan of the paper is as follows. In section \ref{1dhm}, we derive a 4D Lorentz covariant \emph{Relativistic Dynamics Equation}. We also show that equations (\ref{dudtg}) and (\ref{4drde}) are \emph{not} Lorentz covariant. They are, however, canonically embedded in (\ref{uam1}). By taking the tensor $A$ to be constant, we obtain a covariant definition of \emph{uniformly accelerated motion}. In section \ref{explicitsolns}, we obtain explicit solutions to equation (\ref{uam1}) in the case of a constant force. By integrating, we obtain trajectories of a uniformly accelerated point, or, equivalently, a uniformly accelerated observer. We conclude this section by computing the nonrelativistic limits of our solutions. In section \ref{conc}, we discuss the continuation of this research.

Accelerated reference frames were also studied in \cite{Nelson} and \cite{Turyshev}.

\section{Covariant Relativistic Dynamics Equation}\label{1dhm}
$\;\;\;$

In this section, we derive a 4D Lorentz covariant \emph{Relativistic Dynamics Equation} (equation (\ref{uam2}) below). We also show that equations (\ref{dudtg}) and (\ref{4drde}) are \emph{not} Lorentz covariant. They are, however, canonically embedded in (\ref{uam2}). First, we will review some basic notions and establish our notation.

\subsection{Basic Notions}\label{bn}
$\;\;\;$

In flat Minkowski space, the spacetime coordinates of an event are denoted by
$x^{\mu}\;(\mu=0,1,2,3)$, with $x^0=ct$. The inner product is
\begin{equation}\label{Minkip}
x\cdot y=\eta_{\mu\nu}x^{\mu}y^{\nu},
\end{equation}
where $\eta$ is the Minkowski metric $\eta_{\mu\nu}=\operatorname{diag}(1,-1,-1,-1)$. The worldline of a particle is $x(t)=(ct,\mathbf{x}(t))$. The particle's 3D velocity is  $\mathbf{v}=\frac{d\mathbf{x}}{dt}$. Then $\frac{dx}{dt}=(c,\mathbf{v})$, and the dimensionless scalar $\gamma$ is defined by
\begin{equation}\label{defgamma}
\gamma=\gamma(\mathbf{v})=\frac{1}{\left|\frac{dx}{cdt}\right|}=\frac{1}{\sqrt{1-\frac{\mathbf{v}^2}{c^2}}}.
\end{equation}
The particle's \textit{proper time}, denoted by $\tau$, is defined by
\begin{equation}\label{defproptime}
\gamma d\tau=dt.
\end{equation}
Since $cd\tau=ds$, where $ds$ is the differential of arc length along the particle's worldline, the proper time is a Lorentz invariant quantity.
The particle's dimensionless \textit{four-velocity} $u^{\mu}$ is defined, as usual, by
\begin{equation}\label{4vel}
u^{\mu}=\frac{dx^{\mu}}{ds}=\frac{1}{c}\frac{dx^{\mu}}{d\tau}, \quad (u^0,u^1,u^2,u^3)=\gamma\left(1,\frac{\mathbf{v}}{c}\right),
\end{equation}
and its \textit{proper velocity} $\mathbf{u}$ is $c$ times the spatial part of the four-velocity:
\begin{equation}\label{propvel}
\mathbf{u}=\gamma(\mathbf{v})\mathbf{v}.
\end{equation}
A straightforward calculation shows that we can write $\gamma$ as a function of the proper velocity:
\begin{equation}\label{gammau}
\gamma=\sqrt{1+\frac{\mathbf{u}^2}{c^2}}.
\end{equation}
The four-velocity always has ``length" $1$ in the Minkowski metric:
\begin{equation}\label{leng}
|u|=\sqrt{\eta_{\mu\nu}u^{\mu}u^{\nu}}=1.
\end{equation}
The particle's \textit{four-acceleration} $a^{\mu}$ is defined by
\begin{equation}\label{4a}
a^{\mu}=c\frac{du^{\mu}}{d\tau}
\end{equation}
and has units of acceleration. Differentiating $u \cdot u = 1$, we see that the four-acceleration and the four-velocity are always perpendicular:
\begin{equation}\label{4perp}
u\cdot a=\eta_{\mu\nu}u^{\mu}a^{\nu}=0.
\end{equation}
This implies that the four-acceleration is spacelike.

The \emph{rest-mass} of an object is denoted by $m_0$, and we let $m=m(\mathbf{v})=m_0\gamma=m_0\gamma(\mathbf{v})$. The \emph{3D momentum} is $\mathbf{p}=m\mathbf{v}=m_0\gamma\mathbf{v}=m_0\mathbf{u}$, and the \emph{four-momentum} is $p=m_0cu=(m_0\gamma c,m_0\gamma\mathbf{v})=(mc,\mathbf{p})$.

\subsection{Embedding $\mathbf{F}=\frac{d\mathbf{p}}{dt}$ in Four Dimensions}\label{4Dcrde}
$\;\;\;$
The standard \emph{Relativistic Dynamics Equation} is the 3D equation
\begin{equation}\label{dudtg2}
\mathbf{F}=\frac{d\mathbf{p}}{dt}.
\end{equation}
In special relativity, however, we require a 4D version of this equation. Since the 3D vector $\mathbf{p}$ is part of the four-momentum $p$, we seek an appropriate expression for $\frac{dp}{d\tau}$ (or $\frac{du}{d\tau}$).

It is natural to consider the 4D equation
\begin{equation}\label{4drde2}
F=\frac{dp}{d\tau}.
\end{equation}
This equation, however, \emph{has no solution} when $F$ is a \emph{constant} four-vector. To see this, suppose $F$ is constant. Then, since $F \sim a$, equation (\ref{4perp}) implies that $a$ is both lightlike and perpendicular to the timelike vector $u$, which is impossible. This implies that the four-acceleration \textit{cannot} be constant in an inertial frame. Hence, equation (\ref{4drde2}) cannot be used to model constant-force motion and is inappropriate as a dynamics equation.

Next, we show that $\mathbf{F}=\frac{d\mathbf{p}}{dt}=m_0\frac{d\mathbf{u}}{dt}$ can be written in the form $c\frac{du}{d\tau}=Au$, where $A$ is an antisymmetric tensor. Since $u=(\gamma,\mathbf{u}/c)$, we have
\begin{equation}\label{m0dudt}
\frac{du}{d\tau}=\left(\frac{d\gamma}{d\tau},\frac{d\mathbf{u}}{cd\tau}\right).
\end{equation}
Using (\ref{gammau}) and then (\ref{defproptime}), we have
\begin{equation}\label{dmdtau}
\frac{d\gamma}{d\tau}=\frac{d\gamma}{d\mathbf{u}}\frac{d\mathbf{u}}{d\tau}=\frac{\mathbf{u}/c^2}{\gamma}\cdot \gamma \frac{d\mathbf{u}}{dt}=\frac{\mathbf{u}}{c^2}\cdot \frac{d\mathbf{u}}{dt}=\frac{1}{m_0c^2}\mathbf{u}\cdot\mathbf{F},
\end{equation}
and
\begin{equation}\label{dpdtau}
\frac{d\mathbf{u}}{d\tau}=\gamma\frac{d\mathbf{u}}{dt}=\frac{\gamma}{m_0}\mathbf{F}.
\end{equation}
Combining (\ref{dmdtau}) and (\ref{dpdtau}), we have
\begin{equation}\label{rdematrix}
c\frac{du}{d\tau}
=\frac{1}{m_0}\left(\begin{array}{ll}0 & \mathbf{F}^T\\ \mathbf{F} & 0 \end{array}  \right)u,
\end{equation}
where the superscript $T$ denotes matrix transposition. This shows that the 3D Relativistic Dynamics Equation (\ref{dudtg2}) is equivalent to
\begin{equation}\label{embiff}
c\frac{du}{d\tau}=Au,\;\;\mbox{ with } \;\;A=\frac{1}{m_0}\left(\begin{array}{ll}0 & \mathbf{F}^T\\ \mathbf{F} & 0 \end{array}  \right).
\end{equation}
Note that $A$ is a tensor of the particular form $\left(\begin{array}{ll}0 & \mathbf{g}^T\\ \mathbf{g} & 0 \end{array}  \right)$, where $\mathbf{g}=\frac{1}{m_0}\mathbf{F}$. As an operator, $A=A^\mu_\nu$ has mixed indices, one upper and one lower. If we lower the upper index using the Minkowski metric, $A_{\mu\nu}$ is antisymmetric.

\subsection{Achieving Lorentz Covariance}\label{canemb}
$\;\;\;$
A tensor of the form $\left(\begin{array}{ll}0 & \mathbf{g}^T\\ \mathbf{g} & 0 \end{array}  \right)$ is not Lorentz covariant. In fact, a Lorentz transformation of such a tensor will, in general, produce \emph{any} antisymmetric tensor. Therefore, in order to achieve Lorentz covariance, we must allow $A$ to be \emph{any} antisymmetric tensor. In fact, if $F=Au$, where is a tensor, then $A$ \emph{must} be antisymmetric. To see this, first note that since, from (\ref{4drde2}), $F\sim a$, we have, from (\ref{4perp}), that $u \cdot F = 0$. Substituting $F=Au$, we obtain
\[ 0
=\eta_{\mu\nu}u^{\mu}F^{\nu}=\eta_{\mu\nu}u^{\mu}A^{\nu}_{\alpha}u^{\alpha}
=u^{\mu}A_{\mu\alpha} u^{\alpha},\] and so
\begin{equation}\label{antisy}
A_{\alpha\beta}=-A_{\beta\alpha}.
\end{equation}
The need for antisymmetry can be understood as follows. The Lorentz transformation produces a \emph{rotation} in Minkowski spacetime. Similarly, \textit{acceleration} can be interpreted as a rotation of the four-velocity, since the four-acceleration is perpendicular to the four-velocity. It is known that a rotation in 3D Euclidean space is given by the exponent of an antisymmetric tensor. The antisymmetry of $A$ is the 4D extension of this fact.

We thus arrive at a
\emph{4D Lorentz covariant Relativistic Dynamics Equation}
\begin{equation}\label{uam2}
\boxed{c\frac{du^{\mu}}{d\tau}=A^{\mu}_{\nu}u^{\nu}}
\end{equation}
\vskip0.2cm\noindent
where $A_{\mu\nu}$ is an antisymmetric rank 2 tensor, or, equivalently, $A^\mu_\nu$ is skew adjoint with respect to the inner product (\ref{Minkip}). The components of $A$ have units of acceleration and may be \emph{functions} of the position $x$ and the four-velocity $u$. We refer to $A$ as the \emph{acceleration tensor} associated with the given motion. Equation (\ref{uam2}) solves the first problem mentioned in the introduction.

In the $1+3$ decomposition, the tensors $A_{\mu\nu}$ and $A_\nu^\mu$ take the form
\begin{equation}\label{aab}
A_{\mu\nu}(\mathbf{g},\boldsymbol{\omega})=\left(\begin{array}{cc}0 & \mathbf{g}^T\\ &
\\-\mathbf{g}&-c\pi(\boldsymbol{\omega})\end{array}\right),\quad
A_\nu^\mu(\mathbf{g},\boldsymbol{\omega})=\left(\begin{array}{cc}0 & \mathbf{g}^T\\ &
\\\mathbf{g}&c\pi(\boldsymbol{\omega})\end{array}\right),
\end{equation}
where $\mathbf{g}$ is a 3D vector with units of acceleration, $\boldsymbol{\omega}$ is a 3D vector with units of $1/\hbox{time}$, and, for any 3D vector $\boldsymbol{\omega}=(\omega^1,\omega^2,\omega^3)$,
\[ \pi(\boldsymbol{\omega})= \varepsilon_{ijk}\omega^k, \]
where $\varepsilon_{ijk}$ is the Levi-Civita tensor. The factor $c$ in $A$ provides the necessary units of acceleration. The 3D vectors $\mathbf{g}$ and $\boldsymbol{\omega}$ are related to the linear, or translational, acceleration and the angular velocity, respectively, of the motion. We will obtain a more precise explanation of the physical meaning of these vectors in \cite{FS2}. We will also show there that if the uniformly accelerated system was at rest at time $t=0$, then $\mathbf{F}/m_0$ is the constant acceleration in the comoving frame.

By (\ref{embiff}), $\mathbf{F}=\frac{d\mathbf{p}}{dt}$ is equivalent to a 4D equation
\begin{equation}\label{auspat}
c\frac{du}{d\tau}=\left(\begin{array}{cc}0 & \mathbf{g}^T\\ &
\\\mathbf{g}&c\pi(\boldsymbol{\omega})\end{array}\right)u,
\end{equation}
where $\boldsymbol{\omega}=0$. Hence, $\mathbf{F}=\frac{d\mathbf{p}}{dt}$ is covariant
only with respect to transformations which preserve the condition $\boldsymbol{\omega}=0$. A straightforward calculation shows that the only Lorentz transformations which preserve the condition $\boldsymbol{\omega}=0$ are boosts in the direction of $\mathbf{F}$ and spatial rotations about the axis of $\mathbf{F}$ . The little Lorentz group, as defined in \cite{Wigner}, is the stabilizer of the spatial axis in a given direction, which we may choose to be the direction of the force $\mathbf{F}$. Thus, $\mathbf{F}=\frac{d\mathbf{p}}{dt}$ is covariant only with respect to this little Lorentz group and \emph{not} to the full Lorentz group. It follows immediately that \emph{1D hyperbolic motion} is also covariant only with respect to this little Lorentz group.

For an additional proof, note that the spatial part of (\ref{auspat}) is
\begin{equation}\label{3dpart}
\frac{d\mathbf{u}}{d\tau} = \gamma\mathbf{g} + \mathbf{u}\times \boldsymbol{\omega},\;\;\hbox{or}\;\; \frac{d\mathbf{p}}{dt} = \mathbf{F} + \gamma^{-1}\mathbf{p}\times \boldsymbol{\omega}.
\end{equation}
Hence, $\mathbf{F}=\frac{d\mathbf{p}}{dt}$ if and only if $\boldsymbol{\omega}=0$.

By taking the tensor $A$ to have \emph{constant} components, we obtain a covariant definition of uniformly accelerated motion, thus solving the second problem mentioned in the introduction. We define \emph{uniformly accelerated motion} as motion whose four-velocity $u(\tau)$ is a
solution to the initial value problem
\begin{equation}\label{uam2ivp}
c\frac{du^{\mu}}{d\tau}=A^{\mu}_{\nu}u^{\nu}\quad , \quad u(0)=u_0,
\end{equation}
where $A_{\mu\nu}$ is an antisymmetric rank 2 tensor with \emph{constant} components, or, equivalently, $A^\mu_\nu$ is constant and skew adjoint with respect to the inner product (\ref{Minkip}). In the next section, we compute \emph{explicit} solutions to (\ref{uam2ivp}).

\section{Explicit Trajectories for Uniformly Accelerated Motion}\label{explicitsolns}

$\;\;\;$
In this section, we obtain explicit trajectories for uniformly accelerated motion, that is, we obtain the explicit solutions $u(\tau)$ of (\ref{uam2ivp}). It is known (see \cite{CO}, page \textbf{1}-65) that for a given initial condition $u(0)=u_0$, (\ref{uam2}) has the unique solution
\begin{equation}\label{exponent solution}
    u(\tau)=\exp (A\tau/c)u_0=\left( \sum_{n=0}^{\infty}\frac{A^n}{n!c^n}\tau ^n\right)u_0\,.
\end{equation}
The worldline $\widehat{x}(\tau)$ of a uniformly accelerated observer may then be obtained by integrating $u(\tau)$.

Since $A$ is antisymmetric, all solutions of the form (\ref{exponent solution}) are Lorentz transformations of the initial velocity $u_0$, with an angle that is linear in $\tau$.

\subsection{Lorentz-Invariant Classification of Solutions}\label{invclass}

Fix $A=A^\mu_\nu(\mathbf{g},\boldsymbol{\omega})$ as in (\ref{aab}). It can be shown by direct calculation that
\begin{equation}\label{detaminuslambdai}
\det(A-\lambda I)=\lambda^4-l_1\lambda^2-c^2(l_2)^2,
\end{equation}
where $l_1=\mathbf{g}^2-c^2\boldsymbol{\omega}^2$ and $l_2=\mathbf{g}\cdot\boldsymbol{\omega}$ are Lorentz invariants, similar to the two known Lorentz invariants associated with the electromagnetic field. Hence,
the matrix $A^\mu_\nu(\mathbf{g},\boldsymbol{\omega})$ has the eigenvalues $\pm\alpha$ and $\pm i\beta$, where
\[ \alpha=\sqrt{\frac{\sqrt{(l_1)^2+4(cl_2)^2}+l_1}{2}}\quad\hbox{ and }\quad \beta=\sqrt{\frac{\sqrt{(l_1)^2+4(cl_2)^2}-l_1}{2}}.\]

We classify our solutions for uniformly accelerated motion into four types, depending on the values of the Lorentz invariants and the eigenvalues of $A^\mu_\nu(\mathbf{g},\boldsymbol{\omega})$:
\[ \begin{array}{ccccc}
\hbox{\underline{Type}} & \; &\hbox{\underline{Lorentz invariants}} &\;& \hbox{\underline{Eigenvalues}} \\
   \; &\;    &\;&\;&\;\\
 \hbox{Null} &\;& l_1=l_2=0 & \; & \alpha=0,\beta=0\\
     \;& \; &\;&\;&\;\\
\hbox{Linear}& \;&l_1>0, l_2=0 & \; & \alpha = \sqrt{l_1} > 0,\beta=0\\
  \; &\;&\;&\;&\;\\
 \hbox{Rotational}& \;&  l_1<0, l_2=0&\; & \alpha=0,\beta =\sqrt{-l_1} > 0\\
     \; &\;&\;&\;&\;\\
 \hbox{General}&\;& l_2\neq 0 &\; & \alpha > 0,\beta > 0
      \end{array}   \]
Note that each type is a \emph{Lorentz-invariant subset}.
\vskip0.3cm
For each of the four types of uniformly accelerated motion, we now obtain the explicit solutions $u(\tau)$ of (\ref{uam2}).

\subsection{Null Acceleration\quad ($\alpha=0,\beta=0$)}

In this case, $|\mathbf{g}|=c|\boldsymbol{\omega}|$ and $\mathbf{g}\perp \boldsymbol{\omega}$. Direct calculation shows that $A^3=0$. Thus, from (\ref{exponent solution}), we have
\begin{equation}\label{utype0}
\boxed{u(\tau)=u(0)+Au(0)\tau/c+\frac{1}{2}A^2u(0)\tau^2/c^2}
\end{equation}
The four-acceleration is
\begin{equation}\label{atype0}
a(\tau)=c\frac{du}{d\tau}= Au(0) + A^2u(0)\tau/c.
\end{equation}
Despite the apparent dependence of the four-acceleration on $\tau$, we will show in \cite{FS2} that the length of $a(\tau)$ is, in fact, constant.

The worldline $\widehat{x}(\tau)$ of a uniformly accelerated observer is, then, \emph{cubic} in $\tau$ and may be obtained by integrating $u(\tau)$:
\begin{equation}\label{xtype0}
\widehat{x}(\tau)= \widehat{x}(0)+c\int_0^{\tau}u(s)ds=\widehat{x}(0) +u(0)c\tau+ \frac{1}{2}Au(0)\tau^2+\frac{1}{6}A^2u(0)\tau^3/c.
\end{equation}
A similar cubic equation was obtained in \cite{Semon} and page 83 of \cite{F04}.

Note that, in general, $A^2\neq 0$, as can be seen from the example
\[A=\left( \begin{array}{rrrr}
0 & 1 & 0 & 0\\1 & 0 & 0 & -1\\0&0&0&0\\0&1&0&0 \end{array}\right)\quad,\quad A^2=\left( \begin{array}{rrrr}
1 & 0 & 0 & -1\\0 & 0 & 0 & 0\\0&0&0&0\\1&0&0&-1 \end{array}\right) .\]

\subsection{Linear, Rotational and General Acceleration}
$\;$
We will obtain the solutions for the remaining three types using the eigenvalues and eigenvectors of $A$. We will need the following two claims:
\vskip0.3cm\noindent
\textbf{Claim 1}\quad Let $A$ be a skew adjoint matrix. Let $\lambda$ be a non-zero eigenvalue of $A$. If $v$ is an eigenvector of $A$ corresponding to $\lambda$, then $v$ is lightlike.
\vskip0.3cm\noindent
To see this, note that
\[ \lambda v^2 = v \cdot \lambda v = v \cdot Av = -Av \cdot v = -\lambda v \cdot v = -\lambda v^2.\]
Since $\lambda \neq 0$, we must have $v^2=0$.  This proves the claim.
\vskip0.3cm\noindent
\textbf{Claim 2}\quad Let $A$ be a skew adjoint matrix. If $v_1$ is an eigenvector corresponding to the eigenvalue $\lambda$, and $v_2$ is an eigenvector corresponding to the eigenvalue $\mu$, where $\lambda \neq -\mu$, then $v_1$ and $v_2$ are orthogonal.
\vskip0.3cm\noindent
To see this, note that
\[ \lambda (v_1 \cdot v_2)=\lambda v_1 \cdot v_2 = A v_1 \cdot  v_2 = v_1 \cdot -A v_2 =v_1 \cdot -\mu v_2 = -\mu (v_1 \cdot v_2). \]
Since $\lambda \neq -\mu$, we must have $v_1 \cdot v_2 =0$. This proves the claim.
\vskip0.3cm\noindent
We will also need the following notation. Let $r$ be a real four-vector, and let $\{E_0,E_1,E_2,E_3\}$ be linearly independent (possibly complex) four-vectors. Let $c_0,c_1,c_2,c_3$ be the unique (possibly complex) numbers such that $r=\sum_{k=0}^{3}c_kE_k$. Let $\Im(z)$ denote the imaginary part of a complex scalar or vector $z$. Define
\begin{equation}\label{defDv}
\begin{array}{cc}
D_0(r)=c_0E_0+c_1E_1,& D_1(r)=c_0E_0-c_1E_1, \\
\;&\;\\
D_2(r)=c_2E_2+c_3E_3, & D_3(r)=-2\Im(c_2E_2). \end{array}
\end{equation}
\subsection{Linear Acceleration\quad ($\alpha>0,\beta=0$)}

In this case, we have $\alpha = \sqrt{\mathbf{g}^2-c^2\boldsymbol{\omega}^2}$ and $\mathbf{g}\perp \boldsymbol{\omega}$.
Let $E_0$ and $E_1$ be eigenvectors corresponding to the eigenvalues $\alpha$ and $-\alpha$, respectively, and let $E_2$ and $E_3$ be linearly independent eigenvectors corresponding to the eigenvalue 0. Note that we may choose all of the eigenvectors to be real. Then
\[ u(\tau)=c_0E_0e^{\alpha\tau/c}+ c_1E_1e^{-\alpha\tau/c}+ c_2E_2 + c_3E_3, \]
where $u(0)=\sum_{k=0}^{3}c_kE_k$. Using (\ref{defDv}), with $D_\mu=D_\mu(u(0))$, we have
\begin{equation}\label{utype1}
\boxed{u(\tau)=D_0\cosh(\alpha\tau/c)+D_1\sinh(\alpha\tau/c)+D_2}
\end{equation}
We claim that $D_0,D_1,D_2$ are mutually orthogonal. By Claim 2, $D_2 \in \operatorname{Span}(E_2,E_3)$ is orthogonal to both $D_0$ and $D_1$, which belong to $\operatorname{Span}(E_0,E_1)$. To show that $D_0\cdot D_1 =0$, note that $E_0=(1/2c_0)(D_0+D_1)$ and $E_1=(1/2c_1)(D_0-D_1)$ are lightlike, by Claim 1. Hence,
$(D_0)^2 \pm 2D_0\cdot D_1 + (D_1)^2=0$, implying that $D_0\cdot D_1 =0$. This also implies that $(D_1)^2=-(D_0)^2$.

From (\ref{4a}) and (\ref{utype1}), we obtain
\begin{equation}\label{atype1}
a(\tau)=\alpha D_0\sinh(\alpha\tau/c)+\alpha D_1\cosh(\alpha\tau/c).
\end{equation}
The length of $a(\tau)$ is constant:
\begin{equation}\label{a2type1}
a^2(\tau)=-\alpha^2 (D_0)^2,
\end{equation}
and thus satisfies (\ref{a2isconst}).
Since $a$ is spacelike, $D_0$ is timelike.

The interpretation of the solutions (\ref{utype1}) are as follows. In the plane generated by $D_0$ and $D_1$, there is 1D hyperbolic motion, and this plane is moving in a normal direction with four-velocity $D_2$. These solutions form a \emph{covariant extension of 1D hyperbolic motion}.

\subsection{Rotational Acceleration\quad ($\alpha=0,\beta>0$)}

In this case, we have $\beta = \sqrt{c^2\boldsymbol{\omega}^2-\mathbf{g}^2}$ and $\mathbf{g}\perp \boldsymbol{\omega}$.
Let $E_2$ and $E_3$ be eigenvectors corresponding to the eigenvalues $i\beta$ and $-i\beta$, respectively, and let $E_0$ and $E_1$ be linearly independent eigenvectors corresponding to the eigenvalue 0.  Since the two complex eigenvalues are complex conjugates of each other, we may choose
$E_3=\overline{E}_2$. Then
\[ u(\tau)=c_0E_0+ c_1E_1+ c_2E_2e^{i\beta\tau/c}+ c_3\overline{E}_2 e^{-i\beta\tau/c}, \]
where $u(0)=\sum_{k=0}^{3}c_kE_k$. Since $u(\tau)$ is real, we must have $c_3=\overline{c_2}$. Using (\ref{defDv}), with $D_\mu=D_\mu(u(0))$, we have
\begin{equation}\label{utype2}
\boxed{u(\tau)=D_0+D_2\cos(\beta\tau/c)+D_3\sin(\beta\tau/c)}
\end{equation}

We claim that $D_0,D_2,D_3$ are mutually orthogonal. By Claim 2, $D_0 \in \operatorname{Span}(E_0,E_1)$ is orthogonal to both $D_2$ and $D_3$, which belong to $\operatorname{Span}(E_2,E_3)$.
To show that $D_2\cdot D_3 =0$, note that $E_2=1/2ic_2(iD_2+D_3),E_3=1/2ic_3(iD_2-D_3)$ are lightlike, by Claim 1. Hence,
$-(D_2)^2 \pm 2iD_2\cdot D_3 + (D_3)^2=0$, implying that $D_2\cdot D_3 =0$. This also implies that $(D_2)^2=(D_3)^2$.

From (\ref{4a}) and (\ref{utype2}), we obtain
\begin{equation}\label{atype2}
a(\tau)=-\beta D_2\sin(\beta\tau/c)+\beta D_3\cos(\beta\tau/c).
\end{equation}
The length of $a(\tau)$ is constant:
\begin{equation}\label{a2type2}
a^2(\tau)=\beta^2 (D_2)^2,
\end{equation}
and thus satisfies (\ref{a2isconst}).
Since $a$ is spacelike, $D_2$ is also spacelike. Substituting $\tau=0$ into (\ref{utype2}), we obtain that $D_0$ is timelike.

The interpretation of the solutions (\ref{utype2}) are as follows. In the plane generated by $D_2$ and $D_3$, there is pure rotational motion, and this plane is moving in a normal direction with four-velocity $D_0$. The solutions (\ref{utype2}) form a \emph{covariant extension of pure rotational motion}.

\subsection{General Acceleration\quad ($\alpha>0,\beta>0$)}

Since the four eigenvalues $\pm\alpha,\pm i\beta$ are distinct, there are linearly independent eigenvectors $E_0,E_1,E_2,E_3$ of $\alpha,-\alpha,i\beta,-i\beta$, respectively. Since the two complex eigenvalues are complex conjugates of each other, we may choose
$E_3=\overline{E}_2$. Then
\[ u(\tau)=c_0E_0e^{\alpha\tau/c}+ c_1E_1e^{-\alpha\tau/c}+ c_2E_2e^{i\beta\tau/c}+ c_3\overline{E}_2 e^{-i\beta\tau/c}, \]
where $u(0)=\sum_{k=0}^{3}c_kE_k$.
Since $u(\tau)$ is real, we must have $c_3=\overline{c_2}$. Using (\ref{defDv}), with $D_\mu=D_\mu(u(0))$, we have
\begin{equation}\label{utype4}
\boxed{u(\tau)=D_0\cosh(\alpha\tau/c)+D_1\sinh(\alpha\tau/c)+D_2\cos(\beta\tau/c)+D_3\sin(\beta\tau/c)}
\end{equation}

As in the previous cases, the vectors $D_0,D_1,D_2,D_3$ are mutually orthogonal, $(D_1)^2=-(D_0)^2$, and $(D_2)^2=(D_3)^2$.

From (\ref{4a}) and (\ref{utype4}), we obtain
\begin{equation}\label{t2a}
a(\tau)=\alpha D_0\sinh(\alpha\tau/c)+\alpha D_1\cosh(\alpha\tau/c)-\beta D_2\sin(\beta\tau/c)+\beta D_3\cos(\beta\tau/c).
\end{equation}
The length of $a(\tau)$ is constant:
\begin{equation}\label{t2a2}
a^2(\tau)=-\alpha^2 D_0^2 + \beta^2 D_2^2,
\end{equation}
and thus satisfies (\ref{a2isconst}).

Since, in this case, $\mathbf{g}$ and $\boldsymbol{\omega}$ are not perpendicular, there exists a basis in which they are parallel (see \cite{LL}). Here, we have in fact obtained the explicit form of this basis, namely, $\{D_0,D_1,D_2,D_3\}$. In the plane generated by $D_2$ and $D_3$, there is pure rotational motion, and this plane is uniformly accelerated in a normal direction.

\vskip0.5cm\noindent
This completes all of the cases.
The general solution to (\ref{uam2}) is
\begin{equation}\label{gensolnbytype2}
u(\tau)=\left\{\begin{array}{l}
  u(0)+Au(0)\tau/c+\frac{1}{2}A^2u(0)\tau^2/c^2\;,\;\hbox{if }\alpha=0,\beta=0 \hbox{ (null acceleration)}\\
  \;\\
 D_0\cosh(\alpha\tau/c)+D_1\sinh(\alpha\tau/c)+D_2 \;,\;\hbox{if }\alpha>0,\beta=0 \hbox{ (linear acceleration)}\\
 \;\\
 D_0+D_2\cos(\beta\tau/c)+D_3\sin(\beta\tau/c)\;,\;\hbox{if }\alpha=0,\beta>0 \hbox{ (rotational acceleration)} \\
 \;\\
 D_0\cosh(\alpha\tau/c)+D_1\sinh(\alpha\tau/c)\\
+ D_2\cos(\beta\tau/c)+D_3\sin(\beta\tau/c)\;,\;\hbox{if }\alpha>0,\beta>0 \hbox{ (general acceleration)}
 \end{array} \right\}.
\end{equation}

\subsection{Nonrelativistic Limit}\label{nrl}
$\;\;\;$
Here we compute the nonrelativistic limit ($c \rightarrow \infty$) for uniformly accelerated motion. This limit can be taken in two different ways. The first way is to keep the tensor $A$ constant, which is equivalent to holding the eigenvalues $\alpha$ and $\beta$ constant. The second alternative is to keep the \emph{components} $\mathbf{g}$ and $\boldsymbol{\omega}$ of the $1+3$ decomposition (\ref{aab}) of $A$ constant.

For null acceleration, we must keep the tensor $A$ constant since the condition $\mathbf{g}^2=c^2\boldsymbol{\omega}^2$ does not allow $c$ to vary. Note that the spatial part of $cu$ is the velocity $\mathbf{v}$ in the nonrelativistic limit, and, in particular, the spatial part of $cu(0)$ is $\mathbf{v}(0)$ in this limit. Thus, the nonrelativistic limit of (\ref{utype0}) is
\begin{equation}\label{classicallimitnull}
\mathbf{v}(t)=\lim_{c\rightarrow\infty}\left(\mathbf{v}(0)+A\mathbf{v}(0)\tau/c +O(c^{-2})\right)=\mathbf{v}(0),
\end{equation}
implying that the velocity in such motion is constant. Thus, the nonrelativistic limit of null acceleration is \emph{zero} acceleration. This justifies the name \emph{null} acceleration.

Next, we compute the nonrelativistic limits for linear, rotational, and general acceleration when we keep the tensor $A$ constant. First, we show that the 3D acceleration $\mathbf{a}=\frac{d\mathbf{v}}{dt}$ is the nonrelativistic limit of the spatial part of $a(\tau)$. To see this, note that the spatial part of $a=c\frac{du}{d\tau}$ is $\frac{d\mathbf{u}}{d\tau}$. Using (\ref{dmdtau}), we have
\begin{equation}\label{spparta}
\frac{d\mathbf{u}}{d\tau}=\frac{d}{d\tau}(\gamma\mathbf{v})=\mathbf{v}\frac{d\gamma}{d\tau}+\gamma\frac{d\mathbf{v}}{d\tau}=\left(\frac{\mathbf{u}}{c^2}\cdot \frac{d\mathbf{u}}{dt}\right)\mathbf{v}+\gamma^2\frac{d\mathbf{v}}{dt},
\end{equation}
which tends to $\frac{d\mathbf{v}}{dt}$ in the nonrelativistic limit.

In the linear case, from (\ref{atype1}), we have
\begin{equation}\label{classicallimitlinear}
\lim_{c  \rightarrow \infty}a(\tau)=\alpha D_1.
\end{equation}
Similarly, for the rotational case, from (\ref{atype2}) we get $\lim_{c  \rightarrow \infty}a(\tau)=\beta D_3$, and for general acceleration, from (\ref{t2a}) we get $\lim_{c  \rightarrow \infty}a(\tau)=\alpha D_1+\beta D_3$. Since $D_1$ and $D_3$ do not depend on $c$, the 3D acceleration $\mathbf{a}$ is constant in all of these cases in the nonrelativistic limit.

To identify the acceleration, we take the nonrelativistic limit of the first equation of (\ref{3dpart}) and substitute $t=0$. This gives  $\mathbf{a}=\mathbf{g} + \mathbf{v}(0)\times \boldsymbol{\omega}$. Thus, the 3D velocity in the nonrelativistic limit  is
\begin{equation}\label{classicallimitlinear2}
\mathbf{v}(t)=\mathbf{v}(0)+(\mathbf{g} + \mathbf{v}(0)\times \boldsymbol{\omega})t.
\end{equation}
The nonrelativistic limit is constant \emph{linear} acceleration. For $\boldsymbol{\omega}=0$, we obtain the usual nonrelativistic result. A nonzero value of $\boldsymbol{\omega}$, however, will change both the magnitude and the direction of the acceleration.

Finally, we compute the nonrelativistic limit by keeping the \emph{components} $\mathbf{g}$ and $\boldsymbol{\omega}$ constant. For linear acceleration, we have $\mathbf{g}^2>c^2\boldsymbol{\omega}^2$, and so we cannot let $c \rightarrow\infty$. Thus, we will consider only rotational and general acceleration. Since, in this case, $\alpha,\beta$ and the $D_i$ all vary with $c$, it is easier to pass to the nonrelativistic limit of the first equation of (\ref{3dpart}), which is
\begin{equation}\label{nrl25}
\frac{d\mathbf{v}}{dt} = \mathbf{g} + \mathbf{v}\times \boldsymbol{\omega}.
\end{equation}
Equation (\ref{nrl25}) describes nonrelativistic motion under a Lorentz-type force. If $\mathbf{g}=0$, we obtain rotation with uniform angular velocity $\boldsymbol{\omega}$. If $\boldsymbol{\omega}=0$, we obtain linear acceleration.

\section{Summary and Discussion}\label{conc}
$\;\;\;$
We introduced a  \emph{covariant} Relativistic Dynamics Equation (\ref{uam2}) which  extends covariantly the 3D relativistic dynamic equation $\mathbf{F}=\frac{d\mathbf{p}}{dt}$. We have shown that the \emph{standard} 4D equation $F=\frac{dp}{d\tau}$ is only partially covariant. To achieve full Lorentz covariance, we replaced the four-force $F$ by a rank 2 antisymmetric \emph{tensor} $A_{\mu\nu}$ acting on the four-velocity.

In section \ref{explicitsolns}, we obtained \emph{explicit} solutions to our  equation in the case of constant force. We call the solutions \emph{uniformly accelerated motion}. We have shown that uniformly accelerated motions are divided into four \emph{Lorentz-invariant} types: null, linear, rotational, and general.  For \emph{null} acceleration, the worldline (\ref{xtype0}) is \emph{cubic} in the time. \emph{Linear} acceleration (\ref{utype1}) covariantly extends 1D hyperbolic motion, while \emph{rotational} acceleration (\ref{utype2}) covariantly extends pure rotational motion.
We have shown that if we keep the tensor $A$ constant, the nonrelativistic limit (\ref{classicallimitlinear2}) of our uniformly accelerated motion is motion with constant linear acceleration. A different nonrelativistic limit is obtained for rotational and general uniform acceleration by keeping the components $\mathbf{g}$ and $\boldsymbol{\omega}$ of the tensor $A$ constant. This limit (\ref{nrl25}) describes motion under a Lorentz-type force, which includes uniform rotation.

We have here computed explicit solutions in the case of constant force. It is natural, therefore, to consider the general case, in which the force varies. In \cite{Fho}, the first author considers the one-dimensional case $F=-kx$.

In \cite{FS2}, we show how to extend equation (\ref{uam2}) to a system of equations which covariantly define a ``comoving frame." We then show that all of the solutions (\ref{gensolnbytype2}) have \emph{constant acceleration in the comoving frame}. The construction of the comoving frame is rigorous and uses Generalized Fermi-Walker transport along the worldline of the observer.
Once we have the comoving frame well defined, we derive spacetime transformations from a uniformly accelerated frame to an inertial frame. We have shown \cite{FG10} that these transformations must be one of two types. Type I assume's Mashhoon's Weak Hypothesis of Locality, which is an extension of the Clock Hypothesis. In Type II, there is a universal maximal acceleration $a_{max}$. In \cite{FS2} we compute explicit Type I transformations.
We also obtain velocity and acceleration transformations as well as time dilation under the Weak Hypothesis of Locality.
%

\vskip0.4cm\noindent
The authors would like to thank B. Mashhoon, F. Hehl, Y. Itin, S. Lyle, {\O}. Gr{\o}n and the referees for challenging remarks which have helped to clarify some of the ideas presented here.

\end{document}